\documentclass[2pt,a4paper]{article}
\usepackage{graphicx}
\usepackage{amsmath}
\usepackage{hyperref}
\title{\bf Higher Order Stability Analysis for Astrophysical Accretion Processes}

\author{\bf Sayan Kundu\thanks{sayan.astronomy@gmail.com} \\
Indian Institute of Technology Indore, \\ Khandwa Rd, Simrol, Madhya Pradesh 453552 \\ \\
\bf Kinsuk Giri\thanks{kinsuk@nitttrkol.ac.in} \\
National Inst. of Technical Teachers‘ Training \& Research \\
Block - FC, Sector - III, SaltLake City, Kolkata - 106, India}
\date{}

\begin{document}
\maketitle
\begin{abstract}
We have done nth order perturbation analysis of the Navier-Stokes equation with the presence of turbulent viscosity in context of thin accretion flow 
around a black hole.  In order to find the stability criteria, we used Green's function
to solve the nth order velocity perturbation equation. 
Finally, we have proposed a velocity stability criterion which helps to find whether the  base flow to be stable for higher order perturbation.

\end{abstract}
\section{Introduction}

Now a days, the study of the stability of a standard thin accretion disk is coming out to be an important topic. Few important issues, viz., how the turbulent viscosity arises, and also the stability criterion due to small perturbation are still a mystery in context of this topic.  
 
Various studies on accretion flows around a black hole have been proposed since last three decades. The stability analysis have been performed for significant part of this proposed accretion disk models.  In case of stability analysis of thin accretion disk it is concluded that the disc is thermally and viscous unstable in the inner region (Lightman \& Eardley 1974, Shakura \& Sunyaev 1976), which excludes the thin disk from the accretion flows in real world.  In addition, the thin accretion flow which is dominated by radiation pressure is found to be  thermally unstable. However, another model for an optically thin, two-temperature accretion flow was put forward by Shapiro et. al. (1976) to explain the hard X-rays from Cyg X-1, and it was found to be thermally unstable (Pringle (1976), Piran (1978)), though this model is been widely studied for X-ray binaries and AGNs (Kusunose and Takahara 1988,1984, Luo and Liang 1994, White and Lightman 1989, Wandel and Liang 1991). Piran (1978) derived a general criteria for thermal and viscous modes for accretion processes. Both optically thin and thick discs with radial advection have been studied extensively (Kato, Honma \& Matsumoto 1988; Abramowicz et al. 1988; Chen et al. 1995; Narayan \& Yi 1994, 1995a, b; Abramowicz et al. 1995; Chen 1995; Nakamura et al. 1996; Misra \& Melia 1996;). If viscous heating could not get balanced by the cooling process via radiation then their has to be a inward energy advection toward the central region and the effect can not be neglected. Narayan et. al. (1995b); Abramowicz et. al. (1995); Chen et. al. (1995) suggested, by analysing the slope of $\frac{\partial M(\Sigma)}{\partial t}$ curve, that both the optically thick and thin advection dominated discs are thermally and viscously stable. Though these stability analysis as well as Piran's criteria are for long wavelength perturbations, a short wavelength perturbations was also been examined by Kato et. al.(1996,1997), and argued that for short wavelength perturbation the ADAF flow is thermally unstable for optically thick discs but thermally stable for the optically thin ones. This result was confirmed by Wu and Li (1996). Again as suggested by Kato (1978) thin disks might get pulsationally unstable. This phenomena was verified by Blumenthal et al (1984) and Wallinder (1990). Stability analysis for accretion flow around a black hole with the effects of Bremsstrahlung and Synchrotron cooling has been studied by Manickam (2004). The stability of accretion flows in presence of nucleosynthesis has been performed by Mukhopadhyay and Chakrabarti (2001). The stability performance for a rotating black hole accretion disk was also been done by Mukhopadhyay (2003).

Here, in this present work we have not tried to get a stability criteria for any special model rather than we tried to get analytically some general stability criterion from the disc performance in higher order perturbation theory. We have taken the flow to be incompressible and have used some approximations for simplicity.

\section{Basic Equations}
We used the fluid equations in Cartesian system and use the Einstein's summation index to simplify the writing. The first one is the continuity equation
\begin{equation}\label{1}
\frac{\partial\rho}{\partial t}+\nabla.(\rho\vec{v})=0
\end{equation}
which we write using summation index as
\begin{equation}\label{2}
\partial_t \rho+\partial_i(\rho v_i)=0
\end{equation}
Here we will consider only the incompressible fluids, so for our purpose the continuity equation becomes,
\begin{equation}\label{14}
\partial_iv_i=0.
\end{equation} 
Next is the Navier-Stokes equation, using the summation index
\begin{equation}\label{3}
\rho(\partial_t v_i+(v_j.\partial_j)v_i)=-\partial_i P+\rho f^{ext}_i+\mu\partial_k\partial^k v_i,
\end{equation}
where $v_i$ is the $i^{th}$ component of the velocity, $\rho$ is the density, $P$ is the pressure and $\mu$ is the coefficient of viscosity. The coefficient of viscosity is dependent on velocity, pressure, density etc.
Here, we are only going to discuss about the Shakura-Sunyeev $\alpha$ prescription (Shakura \& Sunyeev, 1976) where the coefficient of viscosity is taken to be of the form (Giri, 2014)
\begin{equation} \label{4}
\mu=\frac{2}{3}\frac{\alpha a^2 \rho r}{v_{\phi}},
\end{equation}
where, $\alpha$ is a parameter whose value lies between $0$ to $1$, $a$ is the adiabatic sound speed and $v_{\phi}$ is the rotational velocity.

Now, we apply the perturbation series on $\vec{v}, \rho, P, \mu$ as
\begin{equation}\label{5}
A=\sum_{i=0}^{\infty}\lambda_iA^i,
\end{equation}
where, $A$ is the variable on which the perturbation is applied. We did not perturb the gravitational potential due to the Cowling approximation (Cowling, 1941). 

\section{Methodology}
Applying \eqref{5} to \eqref{4} and \eqref{3} and neglecting $\mathcal{O}(\frac{1}{v_\phi^2})$ and higher order terms we get
\begin{equation}\label{6}
\rho^0(\partial_t v_i^k+(v_j^k.\partial_j)v_i^0+(v_j^0.\partial_j)v_i^k)-\mu^0\partial_a\partial^a v_i^k=S_i,
\end{equation}
where, $S_i$ which is source term not including $v_i^k$, $i$ can take value for $x$, $y$ and $z$. Also, $\mu^0$ is given by
\[\mu^0=\frac{2}{3}\frac{\alpha a^2 \rho^0 r}{v_{\phi}^0}\]
and applying \eqref{5} in \eqref{14} for kth order we get,
\begin{equation}\label{15}
\partial_iv_i^k=0.
\end{equation}
We now solve the equation \eqref{6} with the help of Green's function method.  Before that we need to take into account the equation \eqref{15} because it gives a constraint on the fluid flow. We can deal with that condition by observing the term $(v_j^k.\partial_j)v_i^0$ as,
\begin{equation}
(v_j^k.\partial_j)v_i^0
=\partial_j(v_J^k v_i^0)-v_i^0(\partial_jv_j^k)
=\partial_j(v_J^k v_i^0),
\end{equation}
where, the last term goes out because of \eqref{15}. Again we would like to take a look to the term $\partial_j(v_J^k v_i^0)$. We then approximate this term by a constant times $v_i^k$ which is a good  approximation that we assume for the sake of simplicity. Hence, due to this approximation the term reduces 
\[\partial_j(v_J^k v_i^0) \approx Uv_i^k\],  
where, $U$ is a constant, whose value can be inferred as 
\begin{equation}\label{16}
U = \frac{\partial_j(v_J^k v_i^0)}{v_i^k}.
\end{equation}
So our equation of interest becomes,
\begin{equation}\label{7}
\rho^0(\partial_t v_i^k+Uv_i^k+(v_j^0.\partial_j)v_i^k)-\mu^0\partial_a\partial^a v_i^k=S_i.
\end{equation}

Now,  equation \eqref{7} looks like a diffusion equation with an advection part that could be solved by using the Green's function method. Dividing by $\rho^0$ at both sides of  equation \eqref{7} and taking 
both of the Laplace  Fourier transformation of this equation we get 
\begin{equation}\label{8}
G = \frac{1}{-\omega +v^0_iK_i-i\frac{\mu^0}{\rho^0}K^2},
\end{equation}
where, $K_i$ is the component of the wave vector. 

Now, taking the inverse Laplace transform and inverse Fourier transform of \eqref{8}, we have the Green's function as
\begin{equation}\label{9}
G=\frac{A}{\sqrt{4\pi\frac{\mu^0}{\rho^0} t}}e^{\frac{-x^2-y^2-z^2-|\vec{v^0}|^2t^2-Ut^2}{4\frac{\mu^0},{\rho^0}t}}
\end{equation}
where, $A$ is a constant in time.

Now, in case of extremely higher order perturbation, as $S_i$ (in equation \eqref{6}) contains higher order density, pressure etc divided with base density, pressure they could be neglected. The reason can be given as, in the perturbation theory lower order terms dictates the dynamics. So, we can have a homogeneous form of equation \eqref{6} whose solution can be given as 
\begin{equation}\label{10}
v^k_i=\frac{A_i}{\sqrt{4\pi\frac{\mu^0}{\rho^0} t}}e^{\frac{-x^2-y^2-z^2-|\vec{v^0}|^2t^2-Ut^2}{4\frac{\mu^0}{\rho^0}t}}.
\end{equation}

Now, if the flow has to be stable for any time, it's obvious that 
\[|\vec{v^0}|^2+U>0\]
\begin{equation}\label{11}
\implies |\vec{v^0}|^2>-U
\end{equation}

Also, the boundary conditions for accretion disk pose the constraint on the Green's function as 
\begin{equation}\label{12}
v^k_i=\frac{A}{\sqrt{4\pi\frac{\mu^0}{\rho^0} t}}e^{\frac{-r_{min}^2-z^2-|\vec{v^0}|^2t^2-U^it^2}{4\frac{\mu^0}{\rho^0}t}}=0
\end{equation}

at the outer radius $r_{max}$

and
\begin{equation}\label{13}
v^k_i=\frac{A}{\sqrt{4\pi\frac{\mu^0}{\rho^0} t}}e^{\frac{-r_{max}^2-z^2-|\vec{v^0}|^2t^2-U^it^2}{4\frac{\mu^0}{\rho^0}t}}=0
\end{equation}
at the inner radius $r_{min}$.

\section{Results}
 We got a solution for the kth order velocity perturbation for an turbulent accretion disk, equation \eqref{10}. And from that solution we can conclude that if the solution need to be stable at each time
 then the condition in equation \eqref{11} has to be satisfied. If the value of the constant $U$ (\eqref{16}) is positive then we can see the flow stays stable always but if the value of $U$ gets negative then the stability is depended on whether $|\vec{v^0}|^2>U$ or not. If for a flow the constant $U$ becomes zero then also the solution stays stable. So, one can conclude that the term $\partial_j(v_J^k v_i^0)$ really govern the stability for an accretion flow. If we can make it zero then no matter what the flow stays stable for the higher order perturbations.  From equations \eqref{12} and \eqref{13} we can get the instabilities which arises from the boundaries. 

\section{Discussion}
The Green's function solution is performed for the velocity perturbation for an incompressible flow in the Cartesian grid which can also be applied to the local coordinate frame in the accretion disc. The boundary conditions ensure that the global stability must also get satisfied. We have got an approximate stability condition which can be stated for any order of perturbations, for the higher order the source function must become smaller and smaller so that this stability condition must get satisfied for a stable fluid. This condition can be stated on the primary flow and if this get satisfied one can infer that the flow stays stable for all higher order perturbations.  

The solution of Green's function can also be done in the cylindrical polar coordinate system which can give a much more accurate stability condition, along with more appropriate boundary conditions, for both local and global stability. We have a plan to perform this in future. 
 
{}
\end{document}